\documentclass{article}
\usepackage{spconf,amsmath,graphicx}
\usepackage{times}
\usepackage{epsfig}
\usepackage{graphicx}
\usepackage{amsmath}
\usepackage{amssymb}
\usepackage{subcaption}
\usepackage{algorithm,algorithmic}
\usepackage{multirow}
\usepackage{booktabs}
\usepackage{xcolor}
\usepackage{float}
\usepackage{url}
\usepackage{svg}

\def\SS{{\S}}
\usepackage[disable]{todonotes}
\usepackage{bbm}

\newtheorem{theorem}{Theorem}

\title{On Adversarial Robustness of Large-scale Audio Visual Learning}
\name{Juncheng B Li$^{*}$, Shuhui Qu$^*$\thanks{* equal contribution, Accepted to ICASSP 2022}, Xinjian Li, Po-Yao (Bernie) Huang, Florian Metze}
\address{Carnegie Mellon University}
\begin{document}
\ninept
\maketitle
\vspace{-1cm}
\begin{abstract}
As audio-visual systems are being deployed for safety-critical tasks such as surveillance and malicious content filtering, their robustness remains an under-studied area. 
Existing published work on robustness either does not scale to large-scale dataset, or does not deal with multiple modalities.
This work aims to study several key questions related to multi-modal learning through the lens of robustness: 1) Are multi-modal models necessarily more robust than uni-modal models? 2) How to efficiently measure the robustness of multi-modal learning? 3) How to fuse different modalities to achieve a more robust multi-modal model?
To understand the robustness of the multi-modal model in a large-scale setting, we propose a density-based metric, and a convexity metric to efficiently measure the distribution of each modality in high-dimensional latent space. Our work provides a theoretical intuition together with empirical evidence showing how multi-modal fusion affects adversarial robustness through these metrics. 
We further devise a mix-up strategy based on our metrics to improve the robustness of the trained model.
Our experiments on AudioSet~\cite{gemmeke2017audio} and Kinetics-Sounds~\cite{arandjelovic2017look} verify our hypothesis that multi-modal models are not necessarily more robust than their uni-modal counterparts in the face of adversarial examples. We also observe our mix-up trained method could achieve as much protection as traditional adversarial training, offering a computationally cheap alternative.~\footnote{Implementation: https://github.com/lijuncheng16/AudioSetDoneRight}
\end{abstract}
%
%
\vspace{-0.4cm}
\section{Introduction}
\label{sec:intro}
Nowadays, uploading a clip of video or audio to social media platforms such as Facebook or YouTube will trigger a multi-modal content filtering algorithm to proactively search for potentially policy-violating content; home monitoring devices such as NestCams or RingCams are presumably using audio-visual models to identify events in the monitored area.
Multi-modal classification in safety critical, audio-visual tasks therefore calls for a thorough understanding of its robustness, besides its accuracy. 


Many recent researches have documented neural network models could be vulnerable to adversarial attacks, manipulations of the input to a classifier specifically crafted to be inconspicuous to humans, but which cause the classifier to predict incorrectly~\cite{moosavi2017universal,carlini2019evaluating}. 
Concerns about potential adversarial examples have sparked a huge interest in the research community to study how can we train robust models that defend against potential perturbations~\cite{carlini2019evaluating, tsipras2018robustness}. However, building such adversarially robust models is challenging~\cite{tsipras2018robustness}.
A smaller but still substantial line of work has emerged to show that we could have formal verification of the robustness of neural network models~\cite{wong2018provable, jordan2019provable}.
However, such methods are subjected to very tight constraints and are notoriously difficult to scale. So far, despite some successful large-scale empirical evaluations~\cite{cohen2019certified,madry2017towards} on image-only datasets, robustness of multi-modal learning has not been fully understood. As illustrated in Fig.~\ref{fig:multimodal}, the major challenge to analyzing multi-modal models is the high non-convexity and non-linearity of the decision boundaries in high dimensional latent spaces. 

In this work, we discuss robustness of multi-modal neural network models for classification tasks in the large-scale setting. 
We first measure \emph{point-wise robustness} through the empirical maximum allowable perturbation in $\ell_p$ norm. Based on \emph{point-wise robustness}, we show there exist counter examples to the general claim that multi-modal models are more robust compare to the uni-modal models. 
Due to the limitation of point-wise robustness in terms of scalability and generalizability, class-wise robustness is a necessary and practical compliment to tackle large-scale multi-modal robustness problems. 
Instead of measuring the accuracy drop caused by running universal adversarial perturbation in different magnitude and $\ell_p$ norm~\cite{moosavi2017universal}, we define the class-wise metric by using 1) the density of samples within the high-dimensional ball centered at the centroid of each class with a certain $\ell_p$ radius; 2) the convexity of samples in the high dimensional latent space. We evaluate our metric on the AudioSet~\cite{gemmeke2017audio} and Kinetic-Sounds dataset~\cite{arandjelovic2017look}. The results indicate that multi-modal models are only more robust measured by class-wise metrics for a limited number of classes. We also observe the point-wise robustness of classification results vary greatly depending on the variance of the data with specific class label. 

Inspired by our observations, we propose a density-convexity-based mix-up fusion technique to smooth the decision boundary and add robustness to the fused model. Our proposed mixup could both improve class-wise robustness and point-wise robustness upon our baseline fusion model while increasing the accuracy compared to vanilla fusion methods. We also compare it to traditional adversarial training, where we also see competitive robust accuracy. These advances allow us to address adversarial robustness in large-scale multi-modal settings for the first time.

\vspace{-0.3cm}
\section{Related Work}
\label{sec:related}
Previous works such as~\cite{jordan2019provable, wong2018provable, Wu_2020_CVPR} focused on \emph{point-wise} robustness, studying the maximum allowable radius of centered Chebyshev ball: \emph{a $\ell_p$ ball centered at an input point, within which the output class of a given neural network with remains unchanged, treating the decision boundary of the model as a convex or non-convex polytope.} This formulation certainly provide a safe threshold to defend against adversarial attacks, whereas it involves expensive iterative computations on each anchor point, resulting in huge difficulty to scale~\cite{jordan2019provable,wong2018provable}, most of them depend their claims on small-scale datasets like MNIST or CIFAR-10.
For the large-scale multimodal datasets such as AudioSet, such a verification would hardly be feasible.

Some recent works are try to study defence methods in large scale, including adversarial training~\cite{madry2017towards,gan2020large}, randomization~\cite{cohen2019certified}, and model ensemble~\cite{sen2020empir}. Most of them measured robustness by point-wise accuracy or attack success rate for specific attack budget $\epsilon$ and number of iterations. However, these metrics are often too general to reflect the classifier's behavior under the influence of adversarial perturbation. This motivates us to look into both class-wise accuracy changes along with point-wise accuracy change in order to have a better chance of understanding potential reasons of label change caused by adversarial perturbations. 

Audio-visual learning~\cite{baltruvsaitis2018multimodal} itself is more complicated than image classification, and the current focus still seems to be improving accuracy~\cite{wang2020makes}. The robustness of multimodal classification models involving large-scale video-audio dataset has never been studied rigorously. \cite{Wu_2020_CVPR}~considered the robustness of deep neural networks on videos and experimented on UCF101 dataset~\cite{soomro2012ucf101} which is a relatively small dataset. They measured robustness by the maximum safe radius (point-wise), which computes the minimum distance from the optical flow sequence obtained from a given input to that of an adversarial example in the neighbourhood of the input. To our knowledge, our work is the first to comprehensively study the robustness of multi-modals models both consist of video and audio against adversarial perturbation cause changes to both modalities.

\vspace{-0.3cm}
\section{Background}
\subsection{Universal Adversarial Perturbations}
The problem of computing universal adversarial perturbation to attack a classification model $f$ by maximizing the following~\cite{madry2017towards}:


\begin{equation}
\begin{aligned}
&
  \mathbf{E}_{(x,y)\sim \mathcal{D}} \underset{x' \in C(x)}{\max   }[L(f(x'),y)]\\
& \text{subject to  }
C(x)=\{ a\in \mathbb{R} :||a-x||_p \leq \epsilon \}.
 \\
\end{aligned}
\end{equation}

\noindent where $L$ is the loss function, $x$ is input and $y$ is label, $\mathcal{D}$ is the dataset, and $x' = x + \delta$ is our perturbed input. We want to find some perturbation $x'$ that looks ``indistinguishable'' from $x$, yet is classified incorrectly by $f$ even when $x$ is classified correctly.

\todo{Define and reference PGD - otherwise the paper will be hard to understand for the reader}
To solve such a constrained optimization problem, one of the most common methods utilized to circumvent the non-exact-solution issue is the Projected Gradient Descent (PGD) method~\cite{madry2017towards}:
\begin{equation}
    \delta := \mathcal{P}_{\epsilon} \left(\delta - \alpha \frac{\nabla_{\delta}L(f( x +\delta),y)}{\Vert \nabla_{\delta}L(f( x +\delta),y)\Vert_p}\right)
\end{equation}
where $\mathcal{P}_{\epsilon}$ is the projection onto the $\ell_p$ ball of radius $\epsilon$, and $\alpha$ is the gradient step size.

\vspace{-0.3cm}
\subsection{Multi-Modal Adversarial Perturbations}
\label{sec:formulation}

Under audio-visual multimodal learning setting, we formulate our loss as $L_{multi} = L(f (g(x_A) \oplus h(x_V)), y)$, over the classification function $f$, which can however readily be expanded to additional modalities. 
$g(x_A)$ denotes the encoding of audio features into a bottleneck representation with audio encoding networks (CSN) depicted in Figure~\ref{fig:CSN}, while $h(x_V)$ similarly represents the (R2+1D)CNN~\cite{tran2018closer} encoded video representation, and $\oplus$ indicates concatenation of features. $\mathcal{D}_A$ and $\mathcal{D}_V$ indicating their individual dataset. This more complicated setting requires us to study the adversarial perturbation computed against both the audio input $\delta_{A}$ and the video input  $\delta_{V}$, and can be written out as: 

\todo{Reference CSN and 3DCNN here. What is $\delta$ here I thought the input feature is $x$? Is it the perturbation? Yes, it is the perturbation against video and audio}
%
\vspace{-0.3cm}
\begin{equation}
\begin{aligned}
& \mathbf{E}_{(x_A,y)\sim \mathcal{D_A};(x_V,y)\sim \mathcal{D_V} }\underset{\delta_{A} \in C \left( x_{A} \right), \delta_{V} \in C \left( x_{V} \right)}{\max   }
  [L(f(x'),y)]\\
& \text{subject to  }
C(x)=\{a \in \mathbb{R}:||a-x||_p \leq \epsilon \}
\end{aligned}
\end{equation}
Here, $x' = g(x_{A} + \delta_{A}) \oplus h(x_{V}+ \delta_{V})$. The set $C(x)$ is usually defined as a ball of small radius of the perturbation size $\epsilon$ (of either $\ell_\infty, \ell_2 \,$ or $\, \ell_1$) around $x$. 
Thus, to compute uni-modal audio perturbation to attack the multimodal model, our PGD step could be rewritten out as: 
\begin{equation}
\label{eq:audio-loss}
    \delta_{A} :=\mathcal{P}_{\epsilon} \left(\delta_{A} - \alpha \frac{\nabla_{\delta_{A}}L(f(g(x_{A} + \delta_{A}) \oplus h(x_{V})),y)}{\Vert \nabla_{\delta_{A}}L(f(g(x_{A} + \delta_{A}) \oplus h(x_{V})),y)\Vert_p}\right)
\end{equation}
Accordingly, to compute video perturbation against video classifier $g(x)$, we perform the following PGD step:
\begin{equation}
\label{eq:video-loss}
 \delta_{V} := \mathcal{P}_{\epsilon} \left(\delta_{V} - \alpha \frac{\nabla_{\delta_{V}}L(f(h(x_{V} + \delta_{V}) \oplus g(x_{A})),y)}{\Vert \nabla_{\delta_{V}}L(f(h(x_{V} + \delta_{V}) \oplus g(x_{A})),y)\Vert_p}\right)
\end{equation}
In the more complicated multi-modal case, we jointly optimize $\delta_{A}$ and $\delta_{V}$, where: 
\begin{equation}
\begin{aligned}
    \label{eq:multi-loss}
 \delta_{A},\delta_{V} &:= \mathcal{P}_{\epsilon}(\delta_{({V,A})} - 
 \alpha \frac{\nabla_{\delta_{({V,A})}}L(f(h( x_{V} + \delta_{V}) \oplus g(x_{A}+ \delta_{A})),y)}{\Vert \nabla_{\delta_{({V,A})}}L(f(h(x_{V} + \delta_{V}) \oplus g(x_{A} + \delta_{A})),y)\Vert_p})
\end{aligned}
\end{equation}
\vspace{-0.5cm}
\section{Challenge Common Assumptions in Multimodal Learning}

\begin{figure}[t]
    \centering
    \includegraphics[width=0.95\linewidth]{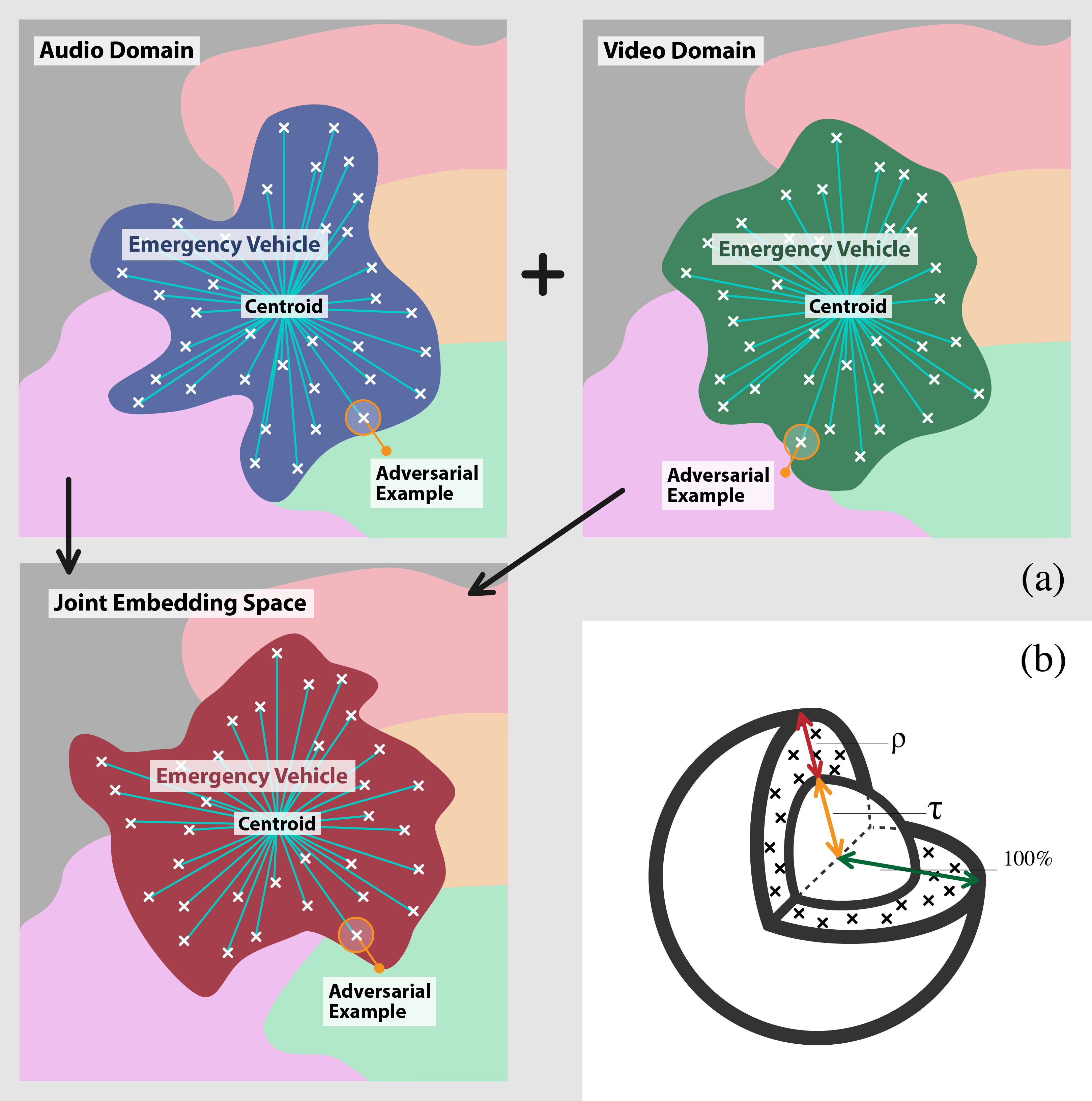}
    \caption{(a) An illustration of multi-modal fusion. (b) Illustration of the centroid based density metric $\rho_c^{R_\tau, p}$.}
    \label{fig:multimodal}
    \vspace{-0.3cm}
\end{figure}

There is a vague notion that multimodal systems are generally more robust compared to unimodal models~\cite{baltruvsaitis2018multimodal}: ``having access to multiple modalities that observe the same phenomenon may allow for more robust predictions.'' 
From an information retrieval perspective, this statement is theoretically true since the same information was captured twice in different modality, improving the robustness of multimodal models. 
However, this is not always true if we rigorously consider how adversarial perturbations affect the neural network model as follows.
\begin{theorem}\label{theo1}
There exists a sample $x_i \in \mathcal{D}$, and a unimodal sample-wise attack $\exists||\delta_{A,i}||_{p} \leq \epsilon_{A}$ or $\exists||\delta_{V,i}||_{p} \leq \epsilon_{V}$  that can break a multimodal fusion network $f((x_{V,i} \oplus x_{A,i}),y_i)$, changing its prediction label $y_i$.
\end{theorem}
\todo{Check that I added the ( in the $f$ correctly}
Here, $\mathcal{D}$ is the dataset, and $\epsilon_{A}$ and $\epsilon_{V}$ are the point-wise robustness threshold for each uni-modal of sample $x_i$. Therefore, as a conjecture, a unimodal attack can break a multimodal model, which we empirically verified the existence of such cases in our experiments. The proof of Theorem~\ref{theo1} can be found in the appendix page.
\todo{Why is $\mathcal{D}$ now a dataset? Previously it was a distribution, at least in the individual modalities?}

\vspace{-0.3cm}
\section{Metrics for Class-wise robustness}

Point-wise robustness is limited in terms of scalability and generalizability. 
Class-wise robustness metric is a more efficient for a large-scale dataset. Instead of exhaustively running universal adversarial perturbation 
we define two metrics to capture the main robustness property of each class.
\vspace{-0.3cm}
\subsection{Centroid-based density metric}
\label{sec:density}
We calculate the class-wise density of the class's high dimensional $l_p$ norm ball by a function of number of samples $n_c$ in the class $c$ and the volume of the $l_p$ norm ball. In this work, $n_c$ is the number of samples of each class in Audioset.

The centroid of a class $\odot_c$ is the mean of bottleneck features $l=g(x)$ of samples $x$ in the class $c$: $\odot_c = \frac{\sum_{i=1}^{n_c}l_{i,c}}{n_c}$,
where $n_c$ is the number of samples in the class $c$. In our case, it is $l=g(x_A)$ for audio modality or $l=g(x_V)$ for video modality. For each class, we calculate the distance of samples in $c$ to the centroid $\odot_c$. The radius of a class $R_{p,c}$ on $l_p$ norm ball is the maximum distance of all samples in $c$ to the centroid $\odot_c$: $R_{p,c} = \max \lVert l_{i,c} - \odot_c \rVert_{p,i=1,...,n_c}$. The radius of first $\tau$ percentage of samples closing to the centroid is $R_{\tau, p, c}$, and the $n_{\tau,c} = \tau \times n_c$ is the number of $\tau$ percentage samples close to the centroid. 
According to~\cite{jorgensen2014volumes}, the volume $V_d^p(R)$ of the $d$-dimensional $l_p$ norm ball with a radius $R$ is: $V_d^p(R) = \frac{(2\Gamma(\frac{1}{p} + 1)R)^2}{\Gamma(\frac{d}{p} + 1)}$, 
where, $d$ is the dimension of the ball, $R$ is the radius, $p$ is the $\ell_p$-norm, and $\Gamma$ is the Gamma function\footnote{\url{https://wikipedia.org/wiki/Gamma_function}}.
\todo{Previously, $\mathcal{D}$ was a dataset, now it is $x$ - can we clarify the notation? Also, what is $\Gamma$ and what is the $\tau$ percentile in the next section? Not sure I follow here.}
Now, we formally define the robustness of a class $c$ with regard to $\tau$ quantile of the class sample $x_c$'s distance to the centroid $\odot_c$ of the class $c$ by:
\begin{footnotesize}
\begin{equation}
    \rho_c^{R_\tau, p,c} = \frac{n_c - n_{\tau,c}}{log(V_d^{p}(R_{p,c})) - log(V_d^{p}(R_{\tau, p,c}))}
\end{equation}
\end{footnotesize}
where the numerator is the number of class samples whose $l_p$ distance to centroid larger than $\tau$ quantile of samples in $c$; $R_{\tau,p,c}$ is the $\tau$ quantile of all class sample's $l_p$ distance to the class's centroid.  We perform the log operation on the volume to reduce the scale of $\Gamma$ function for the ease of computation.
This can be intuitively interpreted as the density in the outer crust of a ball as is shown in Fig.~\ref{fig:multimodal}(b). Generally, the higher the density of the crust, the more robust the samples within/below the crust are.
\subsection{Convexity-based metric}
The convex set $C$ in geometry is defined as a set where given any two points $x_1, x_2 \in C$ in the set, the set contains the whole line segment $x=\theta x_1 + (1-\theta) x_2, \text{with } 0\leq \theta \leq 1$.
Based on our observations and conjecture, we propose the convexity-based metric as one of the robustness measurement of the class. We construct the convex set $S=\{\hat{x}_s | \hat{x}_s=\theta x_1 + (1-\theta) x_2, \theta \sim U[0, 1], \forall x_1, x_2\in C\}$, and sample $n$ points from it $\{ \hat{x}_1, ..., \hat{x}_n | \hat{x}_i \in S \}$. The metric is as follows:
\begin{equation}
    \kappa_c = \frac{\sum^n_{i=1}\mathbbm{1}\{f(\hat{x}_{i}) = y_c\} }{n}
\end{equation}
where $y_c$ is the class label. The higher the $\kappa_c$ is, the more convex the decision boundary of class $c$ is. In this work, we set $n=2000$. Therefore, we use this metric as a proxy to measure how convex the neural network is. We hope to see positive correlation between convexity and robustness.

For both metrics, we use the bottleneck feature $l$ to calculate the value of the metric. In later experiments, we empirically show the effectiveness of our metrics by constrasting with the accuracy drop caused by universal perturbation~\cite{moosavi2017universal}.
\vspace{-0.3cm}
\section{Density-Convexity based Mix-up}
As is noted by~\cite{Lee_2020_CVPR}, mix-up techniques could potentially smoothen the decision boundary via generating virtual training samples by weighted sum of existing training samples, which improves generalizability. Inspired by our density-based and convexity-based metric, we employ a simple adjustment to mixup:
\vspace{-0.1cm}
\begin{footnotesize}
\begin{equation}
\begin{aligned}
    \tilde{x}_A = \alpha x_{Ai} + (1-\alpha)x_{Aj};\\
    \tilde{x}_V = \alpha x_{Vi} + (1-\alpha)x_{Vj};\\
    \tilde{y} = \alpha y_i + (1-\alpha)y_j;
    \end{aligned}
\end{equation}
\end{footnotesize}
\vspace{-0.1cm}
where $\alpha \in [0,1]$,$(x_{Ai},x_{Vi}, y_i)$ and $(x_{Aj} ,x_{Vj}, y_j)$ are two training samples, with both audio and video inputs drawn from 2 different classes $y_i$ and $y_j$, subject to $\kappa_c < T$ ($T$ is an empirical threshold on the convexity) and $\rho_c^{R_\tau, p} >D$ ($D$ is an empirical threshold on the density), for both classes. Both $T$ and $D$ are dataset dependent parameters, $T=0.5, D=8$ in this work. Effectively, we are augmenting the less convex classes of training data with more samples from the ``denser'' samples which are closer to the center of its feature space.
\vspace{-0.6cm}
\section{Experiments}
\subsection{Dataset and Model Setup}
\textit{AudioSet}~\cite{gemmeke2017audio} contains 2~Million 10-second YouTube video clips, summing up to 5,800 hours annotated with 527 types of sound events.
We train and test the models according to the train (1998999 samples) and test split (20126 samples) described in~\cite{vggish}.
The input for the audio branch are matrices of Mel filter bank features. For the visual branch, we employ a (R2+1D)CNN + transformer backbone~\cite{tran2018closer} to encode the spatial-temporal context. The clean performance (with no data augmentation) of our unimodal audio model and audio-visual model are listed in Table~\ref{tab:audioset-fusion} (italic font). \todo{Is Kinetics-Sound really a large-scale corpus?}
\textit{Kinetics-Sounds}~\cite{arandjelovic2017look} is a subset of  Kinetics~\cite{kinetics} that contains 34 classes of audio-related events (22,107 train, 1,504 validation). 
We preprocess the Kinetic-sound dataset in the similar fashion. Our clean multimodal basline: \textit{86.5\%}. We use Kinetic-sounds only for audio-visual performance since its audio-unimodal performance is low and thus not representative.
\todo{Do we reference R(2+1)D network? I will clarify, in this paper, everything is R(2+1)D and it's generally called 3DCNN. Can you clarify which parts of the data we train and test on? We pool AudioSet and Kinetics-Sounds? Or we do something different?}
\begin{figure}[h]
    \centering
    \includegraphics[width=0.98\linewidth]{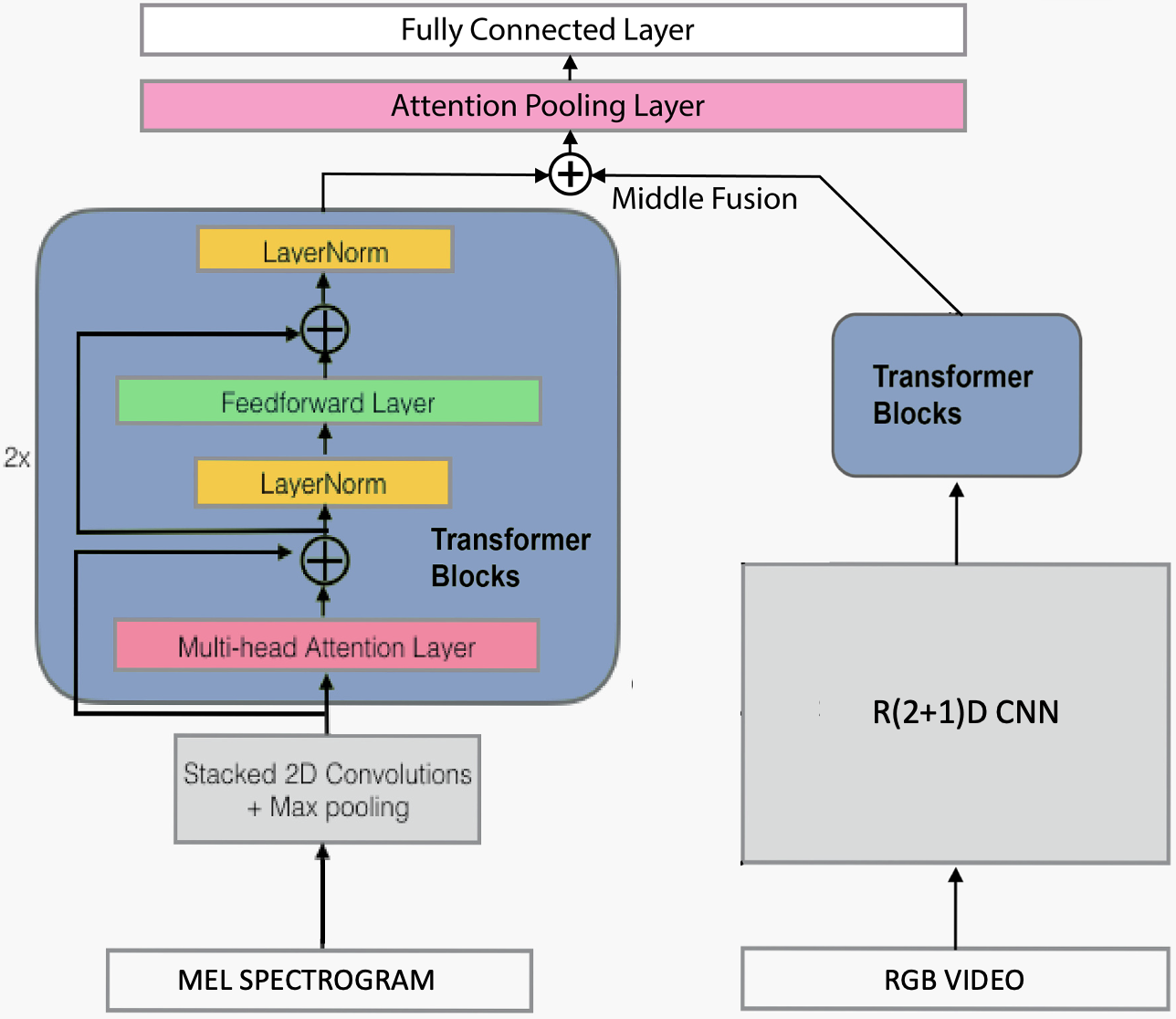}
    \caption{The overall architecture, the audio branch (left) uses Convolution self-attention architecture, video branch is on the right. Mid fusion involves the concatenation step described in \SS\ref{sec:formulation}.}
    \label{fig:CSN}
\end{figure}
\todo{See if the background here can be made transparent, not grey. Reference CSN. Which part of the architecture is CSN? I think it is only the audio part, because this is how we describe it earlier in the paper (Audio: CSN, image: R(2+1)D). But now it is the overall model? Please clarify! Also, here we say that the input to the audio tower is "audio spectrogram input", while elsewhere it is "mel spectrogram". For video, it is "video" - is it RGB, YMCK? Do we have positional encodings only on the audio side?}

\vspace{-0.3cm}
\subsection {Adversarial Perturbation \& Adversarial Training}
All of the adversarial perturbations in this work are computed with the PGD method~\cite{madry2017towards}, with $\epsilon = 0.1$, $\ell_{2}$, trained over 20 iterations. Attack budget ($\epsilon$ and $\#$ of iteration) are constrained to be the same across unimodal, multimodal experiments. Adversarial Training is performed by the methods mentioned in~\cite{wong2020fast}. 

\todo{I think this should be its own subsection and we need to add a bit more detail to it. How were the hyper-parameters of the attack optimized? How were the model training hyper parameters optimized?}

\begin{table}[t]
\setlength\tabcolsep{3.0pt}
\centering
\begin{tabular}{@{}ll | rrr@{}}
\toprule
Models  & Attack   &{\bf mAP} & {\bf AUC} & {\bf d-prime}  \\ \midrule
\textit{Audio UniModal (PANNS)}~\cite{kong2019panns} & No  & 0.383 & 0.963 & 2.521     \\
Audio UniModal  & Yes & \textcolor{red}{0.183}  & 0.895  & 1.770     \\
\textit{Mid Fusion (G-blend)}~\cite{wang2020makes} & No  & \bf{0.427} & 0.971 & 2.686    \\
Mid Fusion     & Yes A+V&\textcolor{red}{0.182} & 0.889   &  1.836 \\
Mid Fusion     & Yes V-only &0.339     & 0.954   &  2.441 \\
Mid Fusion     & Yes A-only &0.310     & 0.940   &  2.276 \\
Mid Fusion mixup & No  & 0.424  & 0.972 & 2.711   \\
Mid Fusion mixup  & Yes A+V&  \textcolor{red}{0.234}   & 0.891   & 1.983  \\
Mid Fusion $AT$ & No  & 0.397  & 0.964 & 2.530   \\
Mid Fusion $AT$ & Yes A+V&  \textcolor{red}{0.199}   & 0.900   & 1.861  \\\bottomrule
\end{tabular}
        \caption{Performance of our best performing CSN models on AudioSet, and their performance against the adversarial perturbation, using the overall architecture shown in Figure~\ref{fig:CSN}. Here, mAP is the mean average precision, AUC is the area under the false positive rate and true positive rate (recall). The d-prime can be calculated from AUC~\cite{gemmeke2017audio}. AT denotes adversarial training.
        A red text color indicates the most potent perturbation against the model.}
        \label{tab:audioset-fusion}
\end{table}

\begin{table}[t]
\centering
\setlength\tabcolsep{3.0pt}
\begin{tabular}{@{}ll | rrr@{}}
\toprule
Models  & Attack &{\bf Acc}  &{\bf mAP} & {\bf AUC}  \\ \midrule
Mid Fusion  & No & \bf{86.5\%} & 0.853 & 0.987    \\
Mid Fusion   & Yes A+V& \textcolor{red}{5.0\%}     & 0.513 &0.824    \\
Mid Fusion   & Yes V-only & 6.6\% &  0.541   &  0.848   \\
Mid Fusion   & Yes A-only & 85.6\% &   0.851 & 0.987     \\
Mid Fusion mixup & No & 85.5\% &  0.854 & 0.987   \\
Mid Fusion mixup  & Yes A+V&\textcolor{red}{15.2\%}  & 0.734  & 0.623     \\\bottomrule
\end{tabular}
        \caption{Performance of CSN models on Kinetics-Sounds.}
        \label{tab:kinetics-fusion}
\end{table}

\subsection{Results and Discussion}
\label{sec:results}

As we can see from Tables~\ref{tab:audioset-fusion} and~\ref{tab:kinetics-fusion}, we find empirical evidence supporting Theorem~\ref{theo1}, as  uni-modal attacks (Video-only, Audio-only) can successfully break the multimodal network. Multimodal attacks are however more potent than unimodal attacks given the same attack budget,\todo{where is the attack budget explained? where do we explain which attack budget we use here, and how this is equal across the two modalities?} with the same $\epsilon$ and $\#$iterations causing more accuracy drop compared to a unimodal attack. 

\begin{figure}[H]
    \begin{subfigure}[h]{0.5\columnwidth}
    \includegraphics[width=1\linewidth]{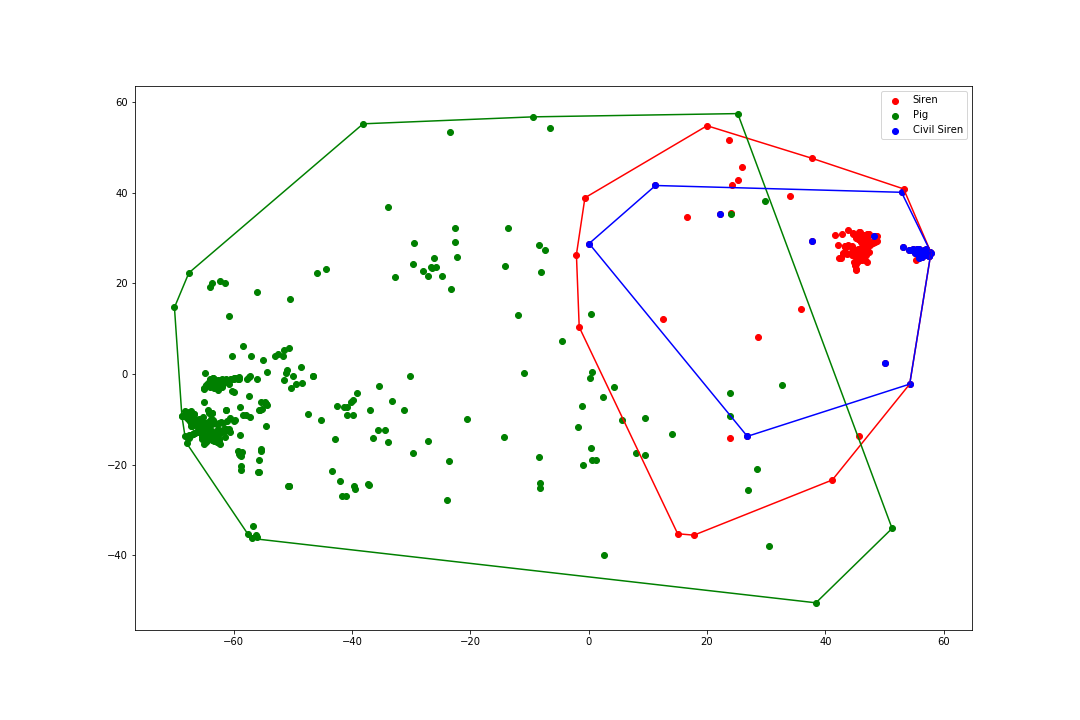}
    \end{subfigure}
    \begin{subfigure}[h]{0.5\columnwidth}
    \includegraphics[width=\linewidth]{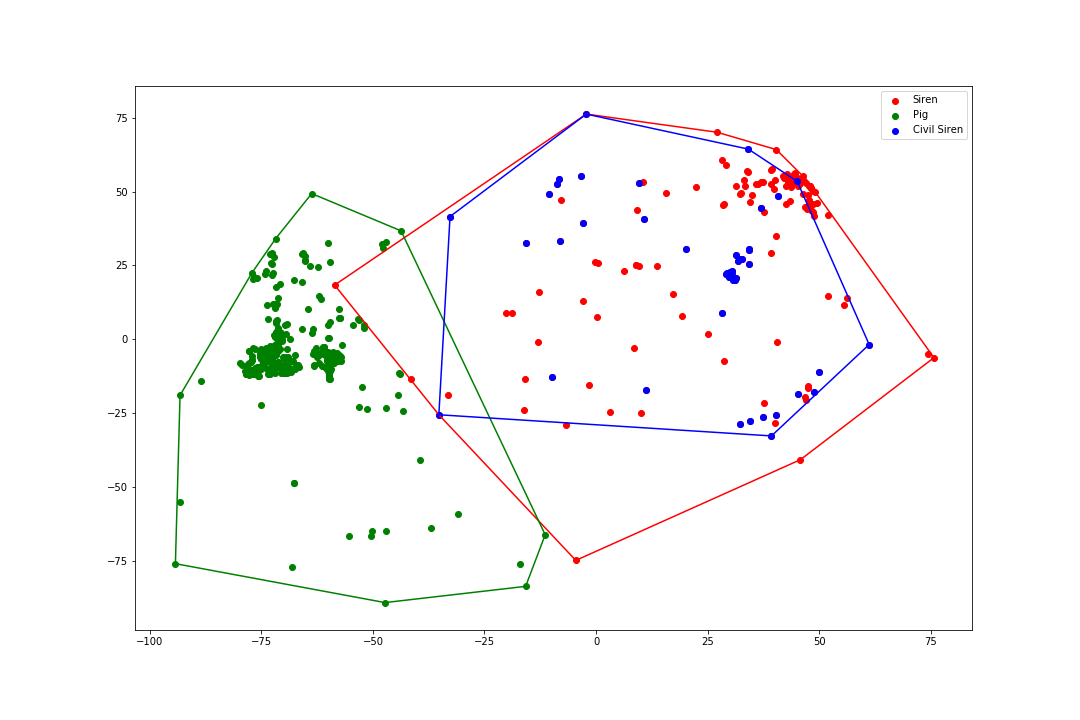}
    \end{subfigure}
    \caption{Convex hull of samples. Left: TSNE of the vanilla fusion feature.  Right: TSNE of the mixup fusion feature. Three classes: Red $\rightarrow$ Siren; Blue $\rightarrow$ Civil Siren (subclass of Siren); Green $\rightarrow$ Pig. } 
    \label{fig:tsne}
\end{figure}

\noindent We could also see the benefit of our proposed mixup, which could help minimize the accuracy loss significantly on both AudioSet and Kinetics-Sounds, compared to vanilla mid fusion and traditional adversarial training (AT). Interestingly, among AudioSet where audio event classification should be the dominant task, audio modality showed on-par robustness compared to video. While in Kinetics-Sounds dataset where video event classification is the dominant task, audio attacks did not affect the classifier at all, video modality dominantly decides the robustness. 

\begin{figure}[H]
    \includegraphics[width=1.2\linewidth]{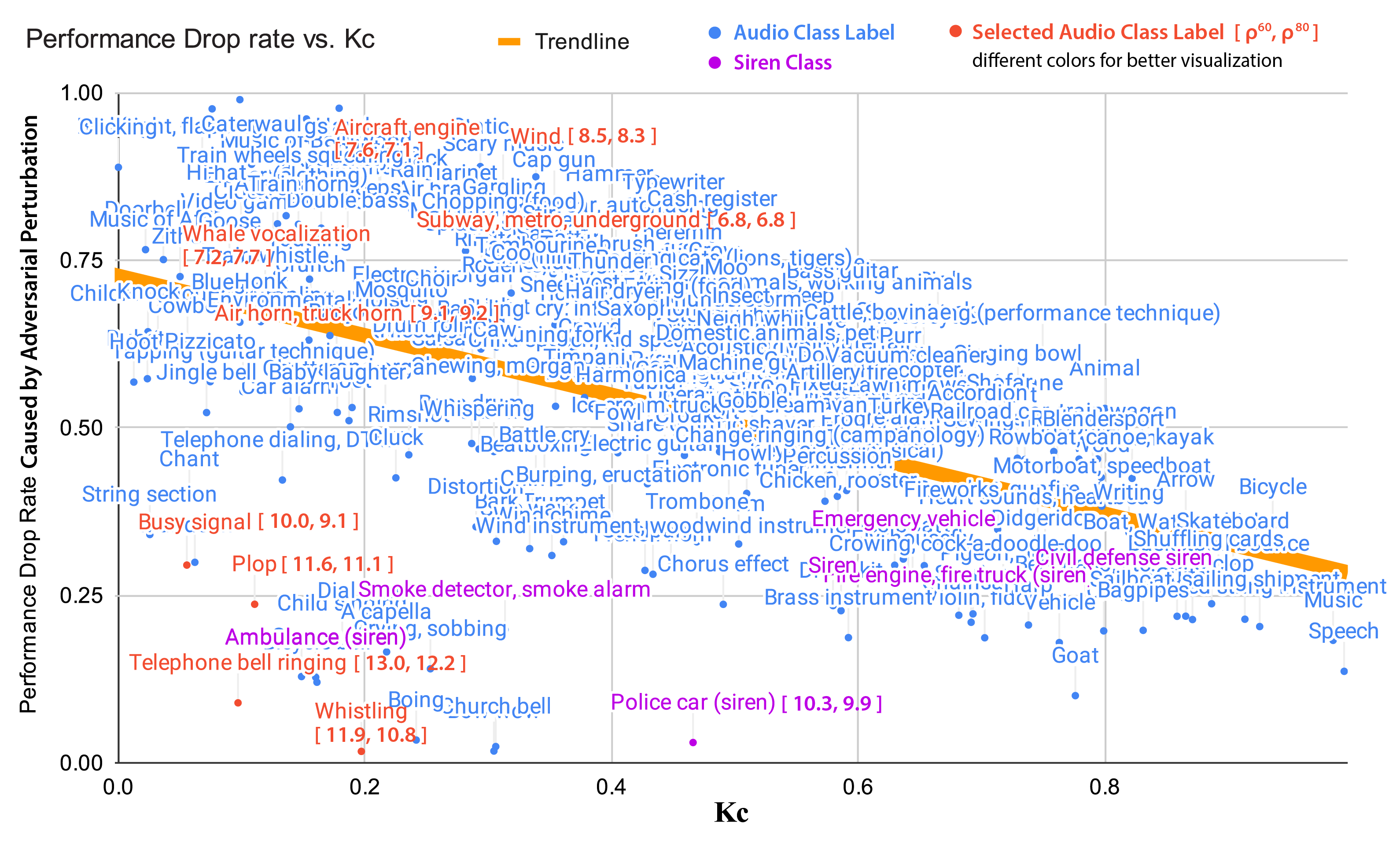}
    \caption{Performance Drop Rate\% VS Convexity ($\kappa_c$) for all audio classes in AudioSet. Density $\rho_c^{R, p}$ partially shown due to space limit.} 
    \label{fig:convexity}
\end{figure}

\noindent In Fig.~\ref{fig:convexity}, Performance Drop Rate$:= \frac{\text{clean performance}-\text{performance under attack}}{\text{clean performance}}$. The density $\rho_c^{R, p}$ (see \SS\ref{sec:density}) is measured at 60 and 80 quantile of the class in $\ell_{2}$ norm. We can observe a negative correlation between drop rate and the convexity of the class, suggesting a positive correlation between robustness and convexity. We could also see higher density would correlate to more robustness for an audio class, some denser classes tend to be more robust despite having low convexity. Intriguingly, we see the pattern that simpler waveforms such as siren (highlighted), bell sounds, child crying are more robust than complex signals like traffic/mechanics sounds. Our empirical findings verify our conjecture that both the density and the convexity of the data are simple factors that could severely affect its robustness. Fig.~\ref{fig:tsne} shows the convex hull of bottleneck feature of mixup fusion model and vanilla mid fusion model, where we observe that the mixup model clearly pushes samples of a class to form a denser outer crust and encourages a clearer boundary between classes resulting in a higher convexity within selected classes.

\vspace{-0.3cm}
\section{Conclusion}

In contrast to some common notion, our work shows that multi-modal systems are not always more robust to adversarial attacks. In this work, we propose alternative metrics to understand the adversarial robustness of large-scale multi-modal models, and we empirically show the effectiveness of our density and convexity metric. We further propose a novel multi-modal mix-up method that selectively augments the denser samples in less convex classes to compensate for the robustness, and which outperforms simple adversarial training with respect to robustness. Our experiments on AudioSet~\cite{gemmeke2017audio} and Kinetic-Sounds dataset~\cite{arandjelovic2017look} verify our hypothesis and the effectiveness of the mix-up strategy.
\todo{I think we need to explain much more clearly how the uni-modal vs multi-modal attack works, and how we make sure that they are comparable, i.e. what the "attack budget" means. Otherwise, reviewers will simply say that if we get to attack both modalities of a multi-modal system, of course we can break it}

\newpage
\section{Acknowledgments}

This work used the Extreme Science and Engineering Discovery Environment (XSEDE), which is supported by National Science Foundation grant number ACI-1548562. Specifically, it used the Bridges-2 system, which is supported by NSF award number ACI-1928147, at the Pittsburgh Supercomputing Center (PSC).
\bibliographystyle{IEEEbib}
\bibliography{egbib}
\appendix

\vfill\pagebreak

\section{Proof of Theorem 1}



Let's consider a binary classification task as an example for simplicity. Let $(x_{A,i}, x_{V,i})$ be a point with different prediction results for audio modality $A$ and video modality $V$. Assume $\exists a,b$, such that $a^T g(x_{A,i}) = -s < 0$ and $b^T h(x_{V,i}) = t > 0$ where $s,t > 0$, and the correct label is $-1$. For the point-wise robustness threshold $\epsilon_{A,i}$ of this point, where an attack $\{\delta_{\epsilon_{A,i}} : ||\delta_{\epsilon_{A,i}}||_P \leq \epsilon_{A,i} \}$ changes the prediction label. By definition, we know $a^Tg(x_{A,i} + \delta_{\epsilon_{A,i}}) \geq 0$ and $a^T g(x_{A,i} + \delta) \leq 0$ for all $0 \leq \delta \leq \epsilon_{A,i}$. If $s<t$, then the fused network predicted the wrong label even without any noise.
\vspace{-0.25cm}
\begin{align}
    f(x_{A,i}, x_{V,i}) &= (a,b)^T (g(x_{A,i}) \oplus h(x_{V,i})) \\
    &= a^T g(x_{A,i}) + b^T h(x_{V,i}) \\
    &= -s + t > 0 
\end{align}
Otherwise, in the case of $s \geq t$,  by applying Intermediate Value Theorem to $g(x)$, there exists a point $0 \leq \delta \leq \epsilon_A$ such that $a^T g(x_{A,i} + \delta) = -t/2$:
\vspace{-0.25cm}
\begin{equation}
\begin{aligned}
        f(x_{A,i}+\delta, x_{V,i}) &= (a,b)^T (g(x_{A,i}+\delta) \oplus h(x_{V,i})) \\
    &= a^T g(x_{A,i}+\delta) + b^T h(x_{V,i}) \\
    &= -\frac{t}{2} + t > 0 
\end{aligned}
\end{equation}

In both cases, we could find a noise $0 \leq \delta < \epsilon_{A,i}$ within the original unimodal robustness threshold to attack the multimodal network successfully. Vise versa for video. Thus, a unimodal attack can break a mulimodal model, which we also empirically verified the existence of such cases in our experiments. (see Table 2 of the main paper). 

We postulate that such phenomenon is like the Mcgurk Effect, where multimodal fusion would further distort the already non-convex decision boundary (Figure~1 of the main paper), making the fused decision boundary very different than the original ones and unpredictable.
\end{document}